\documentclass[a4paper]{article}
\usepackage[T1]{fontenc}
\usepackage[utf8]{inputenc}
\usepackage[italian,english]{babel}
\usepackage[autostyle,italian=guillemets]{csquotes}
\usepackage{multicol}
\usepackage{microtype}
\usepackage{guit}
\usepackage{booktabs}
\usepackage{graphicx}
\usepackage{mathptmx}
\usepackage{amsfonts}
\usepackage{amssymb}
\usepackage{amsmath}
\usepackage{ifpdf}
\usepackage{dcolumn}
\usepackage{booktabs, longtable}
\usepackage{hyperref}
\usepackage{textcomp}
\usepackage{tabularx,array}
\usepackage{ulem}
\usepackage{placeins}
\usepackage{epstopdf}
\usepackage{etex}
\title{New Meteorological and Geological Study \\ of Acquapendente (VT)} 
\author{Tasselli, D. \\ TS Corporation Srl - Astronomy and Astrophysics Department\\ Regione Salamia, 10010 Andrate TO - Italy \\ E-mail:diego.tasselli@tscorporation.org \\ \\ Ricci, S. \\ TS Corporation Srl - Meteorogical and Climatic Change Department\\ Regione Salamia, 10010 Andrate TO - Italy \\ E-mail:stefano.ricci@tscorporation.org \\ \\ Bianchi, P. \\ TS Corporation Srl - Geology and Geophysics Department\\ Regione Salamia, 10010 Andrate TO - Italy \\ E-mail:pamela.bianchi@tscorporation.org}
\date{}
\begin{document}
 
 \maketitle
 
 \begin{abstract} 
In this paper we present an analysis of the geological, meteorological and climatic data recorded in Acquapendente (VT) over 24 years. These data are compared to check local variations,long term trends, and correlation with maen annual temperature. The ultimate goal of this work is to understand logn term climatic changes in this geographic area. The analysis is performed using a statistical approach. From each long series of data calculate the hourly averages and that the monthly averages in order to reduce the fluctuations in the short time due to the day / night cycle. A particular care is used to minimize any effect due to prejudices in case of lack of data. Finally, we calculate the annual average from the monthly ones. 
 \end{abstract}

{\footnotesize This paper was prepared with the \LaTeX \\}
\begin{multicols}%
{2}
\section{\normalsize Introduction} In this paper we present an analysis of geological, meteorological and climatic data releaved in Acquapendente (VT) by ``SMCS - Stazione Meteo-Climatica e Sismologica'' a project by Meteorological and Climatic Change and Geological Department of TS Corporation Srl.\cite{stazionemeteo:2011ug}. \\ We present for the first time an analysis of measurements obtained from local geological \cite{FBarberi:2013} and meteorological data \cite{datimeteo:2013}\cite{eumetsat:2013} and compared in order to check local variations or meteorological conditions. We discuss the geological data, annual temperature means and differences between day time and night time mean values and their comparison with the down time. \subsection{\normalsize Location} Acquapendente is a little town in the center of Italy. Here are identifiable altimeter data, the geographic coordinates and seismic data.\\ \\
\begin{tabular}{c|c}
\hline
\scriptsize \textbf{Latitudine} & \scriptsize \textbf{Longitude} \cr
\hline
\scriptsize $42^\circ 44' 41,28$'' N & \scriptsize $11^\circ 51' 54,36$'' E \cr
\hline 
\end{tabular}  \\  \\ \scriptsize {\bf {Geographic Data of Acquapendente}} \\ \\
\normalsize Acquapendente is located in the extreme north of Latium region near the border with Tuscany region, whose boundaries are marked by a sudden from San Casciano lake, and Umbria region, about six miles north of Bolsena lake at the Natural Reserve of Rufeno's Mount that located on the opposite side of the river Paglia; the town is crossed by the Cassia's street. \\ Acquapendente owes its name to the fact that it is located near several small waterfalls that flow into the river Paglia. \\ The geological map of the town is shown in Geological and Seismic data section. \cite{mappageologica:2013} \\

\begin{tabular}{lp{0.2\textwidth}}
\hline
\scriptsize \textbf{Seismic Zone} & \scriptsize \textbf{Description}  \cr
\hline 
\scriptsize 2B & \scriptsize Area with average seismic hazard may occur where earthquakes strong enough. The sub area 2B indicates a value of ag < 0,20g. \cr
\hline 
\end{tabular} \\ \\ \scriptsize {\bf {Seismic Data of Acquapendente}}
\section{\normalsize Annual data analysis} \normalsize Summers are generally warm and winters are fairly rigid and often winds blow from the north that may last a long time. In winter mornings it is easy to find fog also due to the proximity of the Paglia river. The snow appears about two or three times a year, sometimes in large quantities. The rains dominate the autumn and spring seasons. The following table identifies climate data assigned by Decree of the President of the Republic n. 412 of 26 August 1993.\cite{Tasselli:2011ug}. \\ Characterized by a fairly humid climate, the summer can often record the maximum temperature higher than the district, while in winter with thermal inversion and the particular conditions in which it occurs, it manages to be the coldest city of area, and among the coldest in central Italy (of course compared to the cities in which it is carried out monitoring data). It has a climate for "extreme" in both summer and winter conditions like temperature range and inversion are the order of the day. 
\subsection{\bf \normalsize The temperature inversion}
The temperature inversion, as the name implies, is when the vertical thermal gradient is reversed, ie, when the temperature increases with altitude (positive gradient) instead of decreasing (negative gradient).
Basic conditions so that we can experience nocturnal inversion are essentially three: 
\begin{itemize} 
\item Preparation of morphological country to this phenomenon;
\item Terms of persistent high pressure (skies);
\item Lack of wind.
\end{itemize}
\normalsize The summation of these factors causes a marked loss of heat due to radiation night which leads to the creation of a bearing of cold air that station in the lower layers. \\ \\ This means that the bearing in the early morning minimum Acquapendente registers much lower than the neighboring countries with higher altitude. 
\\ \\
\begin{tabular}{c|c}
\hline
\scriptsize \textbf{Climatic Zone} & \scriptsize \textbf{Day Degrees} \cr
\hline 
\scriptsize E & \scriptsize 2,299 \cr
\hline 
\end{tabular} \\ \\ \scriptsize {\bf {Climatic Parameter of Acquapendente}}
\section{\normalsize Meteo-Climatic Parameter} 
\normalsize In this section we describe air temperatures (T), Dew Point, Humidity, Pressure, Day Time and Night Time Variation, Rain's Days and Fog's Day, obtained by an accurate analysis of the meteorological data from local data by archive \cite{datimeteo:2013}.
We analyzed the parameters given in Table 1. Should be noted that the values considered are related to the last twenty-four-year average and made available for the period 1990-2014. \\ \\
\begin{tabular}{lp{0.10\textwidth}} 
\hline
{\footnotesize Average Annual Temperature}&  {\footnotesize $14,33 ^\circ C$}  \cr
{\footnotesize T average warmest (ago-03)}& {\footnotesize $28,27 ^\circ C$} \cr
{\footnotesize T average coldest (feb-12)}&  {\footnotesize $1,42 ^\circ C$} \cr 
{\footnotesize Annual temperature range}&  {\footnotesize $9,97 ^\circ C$}  \cr
{\footnotesize Months with average T > $20 ^\circ C$} & {\footnotesize  83} \cr
{\footnotesize Total rainfall 1990-2014 [mm]}& {\footnotesize 21401,93} \cr
{\footnotesize Rain Days }&  {\footnotesize 1198}  \cr
{\footnotesize Fog Days }&  {\footnotesize 912}  \cr
{\footnotesize Storm Days} &  {\footnotesize 181}  \cr
{\footnotesize Rain/Storm Days }& {\footnotesize 506}  \cr
{\footnotesize Rain/Snow Days}&  {\footnotesize 21}  \cr
{\footnotesize Rain/Fog Days} & {\footnotesize 159}  \cr
{\footnotesize Rain/Thunder/Fog Days} & {\footnotesize 68}  \cr
{\footnotesize Snow Days} & {\footnotesize 22}  \cr
{\footnotesize Wind Speed max Km/h (Apr-2008)} & {\footnotesize 36,57}  \cr
{\footnotesize Wind Speed min Km/h (Dec-2011)} & {\footnotesize 9,00}  \cr
{\footnotesize Rain max mm (Nov-2008)} & {\footnotesize 285,90}  \cr
{\footnotesize Rain min mm (Jun-2004)} & {\footnotesize 1,50}  \cr
{\footnotesize Earthquake Min (2011/12/20)} & {\footnotesize 0,2 Mw}  \cr
{\footnotesize Earthquake Max (2006/07/06)} & {\footnotesize 3,6 Mw}  \cr
{\footnotesize Earthquake Deep Min (2012/08/21)} & {\footnotesize 0,7 Km}  \cr
{\footnotesize Earthquake Deep Max (2010/04/28)} & {\footnotesize 42,5 Km}  \cr
\hline
\end{tabular} 
\\ \\ \mbox{\bf{\footnotesize  Parameter of this Study}}
\subsection{\normalsize Solar Radiation Territory}  \normalsize The data irradiation of territory taken from the parameters and the data prepared by the European Union, demonstrate the trend of irradiation for the municipality, visible in next table: \\ \\
\begin{tabular}{|c|c|c|c|}
\hline
     \bf {\footnotesize Month} & \bf {\footnotesize DNI}& \bf {\footnotesize Month} & \bf {\footnotesize DNI}\cr \hline
       {\footnotesize Jan} &  {\footnotesize 2410} & {\footnotesize Feb} &  {\footnotesize 3500} \cr
       {\footnotesize Mar} &   {\footnotesize 4030} & {\footnotesize Apr} &  {\footnotesize 4800} \cr
       {\footnotesize May} & {\footnotesize 5570} & {\footnotesize Jun} &  {\footnotesize 6840} \cr
       {\footnotesize Jul} &   {\footnotesize 7990} &  {\footnotesize Aug} & {\footnotesize 7080} \cr
       {\footnotesize Sep} & {\footnotesize 5100} &  {\footnotesize Oct} & {\footnotesize 3710} \cr
      {\footnotesize Nov} & {\footnotesize 2620} & {\footnotesize Dec} &  {\footnotesize 2290} \cr \hline
     \bf {\footnotesize Year} & \bf {\footnotesize 4670} & & \cr
\hline
\end{tabular}
\\ \\ \mbox{\bf{\footnotesize Direct Normal Irradiance (Wh/m$^2$/day) }} \\ \\
\normalsize The weather data and the graphs show extrapolated for the territory covered by the study, including a radiation in the range between 1200 and 1300 kWh /1kWp as map prepared by the European Union \cite{UE:2013} and visible in figure 18, characterized in Over the months, irradiation presented in the graph in Figure 19, which shows the data of the table above, which shows the territory of Acquapendente, a total irradiance Annual of 4670 Wh/m$^2$/day.
\subsection{\normalsize Temperature} \normalsize In this section we describe air temperatures (T) obtained by an accurate analysis of the meteorological data.  The year average temperature has a tendency to remain stable throughout the study period, except: decrease in the period between 1990 and 1996; back and stabilize during 1997-2014. Next graphics show this evidence: \\ \\
\includegraphics[width=0.49\textwidth{}]{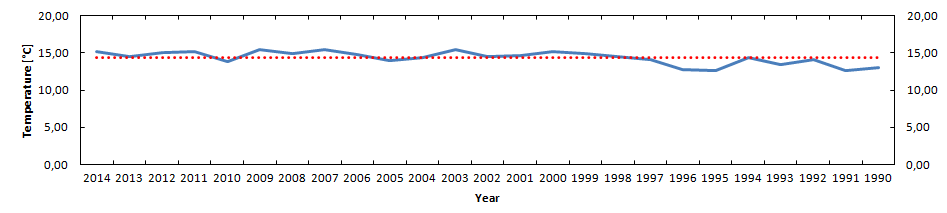} 
{\scriptsize \bf \textit{\textit{{\footnotesize Relationschip of Year Temperature and Average 1990-2014}}}}. \\
 \\The values are calculated considering the entire measurement period (1990-2014), drawing on data from the Annuals published by "Il Meteo.it" \cite{datimeteo:2013} for the period 1990-2014. \\ The study period showed an increase in "heat waves" such as that of 2003, whose negative impact is amplified by the increase in temperature in the urban areas, creating the effect of "heat island" mainly due to soil sealing. \\ \\ \includegraphics[width=0.49\textwidth{}]{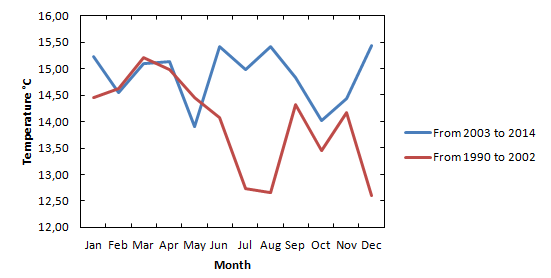} 
 {\scriptsize \bf \textit{\textit{{\footnotesize Variation of Month Temperature about year 1990-2002 and 2003-2014}}}} \\
 \\ The graphs shown in Figures 4 and 5 show the performance of the maximum and minimum temperatures proportional to the average of the period of study (1990-2014). The study shows the following trends:
\begin{itemize}
\item Minimum temperatures: there was an increase of the values recorded with a higher peak in February 2012, with $1,42^\circ C$;
\item Maximum temperatures: evidence for the period 2007-2009 is an increasing trend compared to the 1990-2014 average, while evidence a decrease in the period 1990-1997. The maximum temperature was $28.27^\circ C$ in August 2003;
\end{itemize} 
Table 5 shows the trend of the points of maximum temperature expected in the summer quarter (June-July-August) and are shown in Figure 15. Study shows that the month of August 2003 recorded the highest average maximum value in a context in which the entire month recorded in the period 1990-2014 values always above average.\\ Alway Table 5 shows the trend of the points of minimum temperature expected in the winter quarter (January-February-March) and are shown in Figure 16. Study shows that the month of February recorded the highest average minimum value, in a context in which the entire month recorded in the period 1990-2014 values always above average.
Studying in particular the months with the values of minimum temperature and maximum minors, the following is noted:
\begin{itemize}
\item the month of August (characterized by the presence of the highest value observed in maximum temperatures), shows that trends in temperature has been getting consistently below average for the period 1990 to 2014, while there were two exceedances of this value over the years 2003 and 2012, in agreement with what evident from the graph in Figure 13;
\item the month of February (characterization from the value of the lowest minimum temperature for the period of study), shows that the trend of temperatures has always been, in agreement with what reported in the graph in Figure 14.
\end{itemize}
\subsection{\normalsize Dew point} In this section we describe Dew point obtained by an accurate analysis of the meteorological data. This study showed: 
\begin{itemize}
\item an increase in the year: 1994, 2001, 2002 (the hightest Dew Point period for this study), 2004, 2005, 2006, 2007, 2009, 2010, 2011, 2012 and 2014;
\item a decrese in rest of study period, with the lowest Dew Point in 1991 year. 
\end{itemize}
Graphics in Figure 11 show this trend. 
 \subsection{\normalsize Humidity} In this section we describe humidity obtained by an accurate analysis of the meteorological data. 
 The study highlights a gap in the annual humidity values equal to 67,56\%, calculated according to this formula:
\begin{equation}\frac{av}{am}\end{equation} 
{\footnotesize \textit{ {\bf av} = annual value, {\bf am} = average moisture 1990-2014}} \\ \\ The graph in Figure 9 shows a trend tends to be stable in the values obtained with threshold in growth over the period 1990,1992,1994,1995,1996,2002,2004 and 2010, and a decrease in the period 1991, 1993, 1998,1999, 2000, 2003, 2006, 2007, 2009, 2011 and 2012.
\subsection{\normalsize Pressure} \normalsize In this section we describe pressure obtained by an accurate analysis of the Pressure data. \\ The analysis of the data showed:
\begin{itemize}
\item an increase in atmospheric pressure over the years: 1991, 1998, 2000 (year with the highest level of pressure throughout the study period), 2003, 2004, 2007, 2008, 2011, 2012 and 2014;
\item a decrease in atmospheric pressure over the years: 1993, 1994, 1995, 1996, 1997, 2001, 2002, 2005 (the year with the lowest level of pressure throughout the study period), 2006, 2010 and 2013.
\end{itemize}
In Figure 10 we can see the Pressure Graphics for this study. 
\subsection{\normalsize Day time and night time variation} In this section we describe number of Day time and night time variations obtained by an accurate analysis of the meteorological data. 
The annual averages of the differences between day time and night time temperatures $\Delta T$ have been computed and the results are reported in Table 2. Also in Figure 3, we can see the plot of oscillations of the $\Delta T$ seem to reduce the amplitude during the years. \\ The difference of temperatures, with an average difference of $9,97^\circ C$ (see table 2) does confirm the trend indicate Acquapendente as one of Italian municipalities with higher ratio of the temperature difference between day and night. Table 2 and Figure 3 show these effects.
\subsection{\normalsize Rain's Days} In this section we describe number of Rain's days obtained by an accurate analysis of the meteorological data. \\  An increase in extreme weather events and abnormal, leading to a potential increase in precipitation intensity for each event, especially in areas where there is an increase of average precipitation. \\ \includegraphics[width=0.49\textwidth{}]{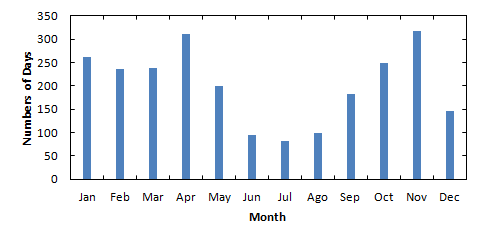} 
{\scriptsize \bf \textit{\textit{{\footnotesize Days Rain for Month from 1990 to 2014}}}}. \\ \\
The increase in intense rainfall of short duration boosts the risk of flash-food, fast flooding of an area circumscribed morphologically, due to the fast ``saturation'' of the surface soil that is no longer able to absorb the rain. The episodes of intense precipitation can also determine the phenomena of surface runoff of rainwater with a possible increase in flooding, but also the risk of water pollution (pollutants from agricultural and road runoff). \\ The study period showed an abnormal increase in rainfall. The pie chart shows the ratio between the total mm of rain measured in a year and the total rainfall measured in the period 1990-2014, according to this formula: \begin{equation}Total Rain=\frac{a}{b}x100\end{equation} 
{\footnotesize \textit{{\bf a} = annual rainfall value, {\bf b} = total value of the rain period}} \\ \includegraphics[width=0.49\textwidth{}]{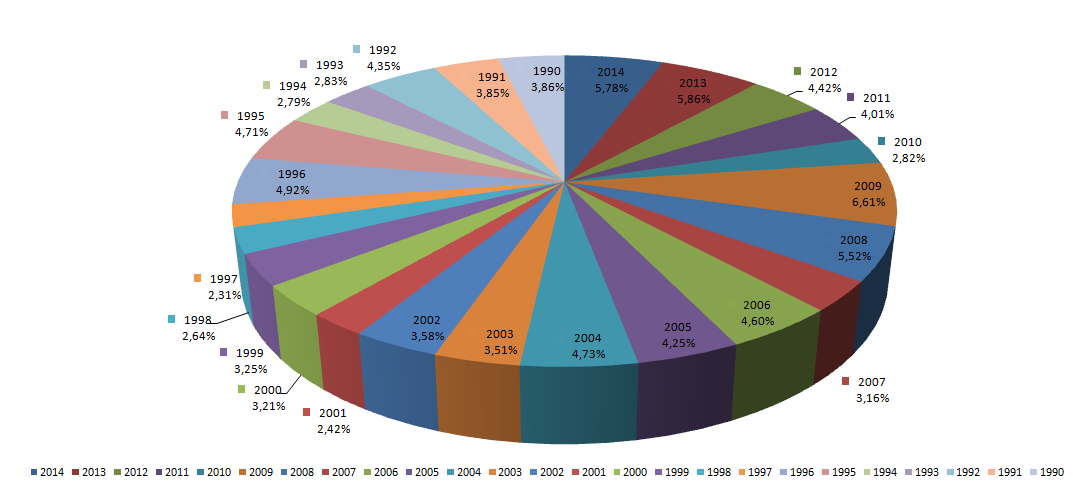}
{\scriptsize \bf \textit{ \% of Rain for this study}}.
\\ \\ In Figure 6 and 7 we can see the total of rain/year in this study. The values are visible in the table 3. Next pie chart evidence type of rain for this study. \\ \\
\includegraphics[width=0.5\textwidth{}]{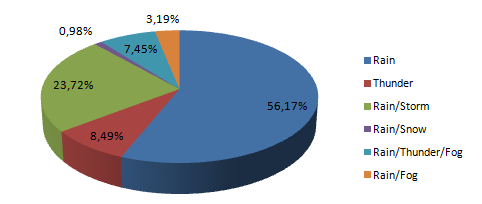}
{\scriptsize \bf \textit{Type of Rain for this study}}. \\ \\ Figure 17 evidence the total rain in mm, for only month. \\ The study shows an abnormal increase in rains, shown in Figure 7, in the years: 2014, 2013, 2012, 2009 (the highest year period mm of rain fell), 2008, 2006, 2005, 2004, 1996, 1995, 1992 , with an average increase of 126.73\% of the total annual rainfall mm which is equal to 856.08 mm on average for year.
\subsection{\normalsize Fog's days} In this section we describe number of Fog's days obtained by an accurate analysis of the meteorological data. \\ The study showed a trend increase in the presence of days with fog and evidence that October is the first month for number of Fog's day and July are the laster. Figure 8 we can see the number of Fog days by this study. \\ \\
\includegraphics[width=0.49\textwidth{}]{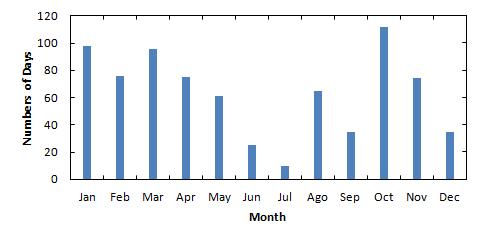}
{\scriptsize \bf \textit{Days Fog for Month from 1990 to 2014}} 
\section{\normalsize Geological and Earthquake data} In this section we describe the Geological and Earthquake data. \\
\includegraphics[width=0.4\textwidth{}]{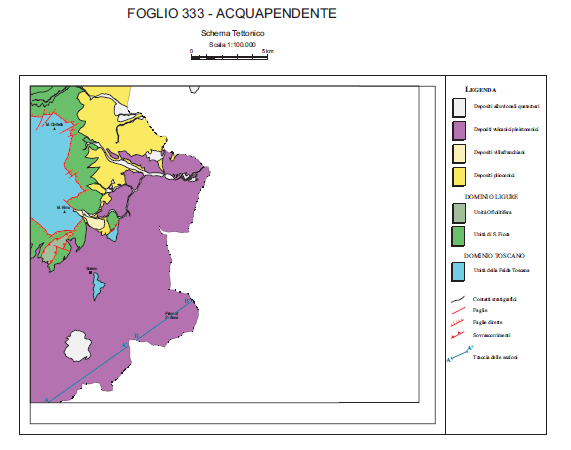} \\
{\scriptsize \bf \textit{Geological Map of Acquapendente - VT }} \cite{mappageologica:2013} \\ \\
The geology in Acquapendente is characterized by high heat flow, locally more than 200 mW$m^{-2}$ \cite{Dellavedova:2013} positive magnetic anomalies \cite{Arisi Rota:2013}, shallow earthquakes, positive gravity anomalies and crustal thinning (20-25 km). The area lies within the catchment area of the river Paglia, and the percentage composition of the various lithologies in the basin is so divided: 37.5\% volcanics, flooding 20\%, 17.5\% flysch, anthropogenic debris cones and 5\%, clay 15\% , sands 2\%. \\ Acquapendente falls in complex lavas and ignimbrites lithoid, and consists of lava flows and ignimbrites lithoid (Pliocene-Pleistocene) interspersed at various levels in the complex pyroclastic. The thickness of this complex is extremely variable and ranges between a few dozen to over a hundred meters. The permeability of the complex is variable and is related to the different degree of cracking and the presence of tectonic reasons. In relative terms the degree of permeability and to be considered medium to high and, together with the complex pyroclastic constitutes the main aquifer area under examination. The field observations combined with the information of the surveys have allowed to estimate more than 2500 m in thickness of the Pliocene deposits \cite{FBarberi:2013}. \\ The predominant lithology consists of bluish clay, interspersed with different stratigraphic levels from debris flow conglomerate and sandstone turbidites. These sediments have been described in detail by \cite{Iaccarino:2013} belong to the first cycle of sedimentation \cite{FBarberi:2013} . \\ Sedimentation during this time period occurred throughout the basin and is also demonstrated by the polls and deep wells for geothermal exploration that have met these sediments beneath the volcanics.. \subsection{\normalsize Description of the main outcrop formations.} The magmatic Pliocene-Pleistocene (rocks apparatus Vulsino) outcropping in the territory of Acquapendente are:
\begin{itemize}
\item \textbf{Basanites and tephrites leucititiche}: The basanites emerge around Mt. Rosso with a dark gray casting very compact. The tephrites leucititiche are present with extensive lava outcropping around Acquapendente. Those visible along the right side of Straw differ from others because of the abundance of phenocrysts of leucite and alteration that turns into a reddish sandy soil.
\item \textbf{Leucitites passers tephrites leucititiche leucititici and basalts}: The leucitites and leucitites passers tephrites rocks are dark gray compact very fine. They are found in lava flows at Acquapendente. The leucitites passers in basalts leucititici alternate, near Mt. Landro, with banks of lapilli and scoria.
\item \textbf{Tuffs yellow pumice gray}: Rest directly on the flysch and the Pliocene as the lava flows and other types of pyroclastic and reach considerable power that exceeds the 50 m. They are characterized by large pumice and black slag of the diameter of 20 - 30 cm. At the base incorporate numerous fragments of lava and limestone marl.
\item \textbf{Tuffs yellow pumice clear}: The pumice tuff yellow light are very compact and generally fill paleo valleys excavated sediments Pliocene tuffs or yellow thick gray pumice. With this acronym indicate levels of tuffs similar to the previous, but with included leucitic, to places with power exceeding 20 m. Surely the various lithological types from different eruptive vents. Typically these horizons are separated by paleo soils that sometimes show phenomena of cooking, which suggests that these levels may have set with phenomena similar to those of ignimbrites.
\item \textbf{Tuffs}: This term indicates an alternation of layers of lapilli, pumice and cineritic separated by paleo soils blacks, each having a thickness varying from a minimum of 10 cm to a maximum of about 2 m. This alternating pyroclastic generally closes a series effusive local.
\end{itemize}
\subsection{\normalsize Spatial structural sub-surface}
\normalsize Based on the gravity data available in the literature it was outlined geometry of the structures that controlled sedimentation and volcanic activity during the Pliocene-Pleistocene area. \\ During the course of the Pleistocene the tectonic structures began to rally the regime undergoing subsidence of the central part. The continental basins linked to the sectors affected by such tectonic regime have developed just before and in conjunction Quaternary volcanic Finally they set important features cross with anti-Apennines trend, visible in next figure. \\ The complex of Vulsini and Sabatini have developed coinciding with the two largest discontinuities observed that interrupt the Pliocene extensional structures and the dorsal midline \\ The transverse structure northernmost coincides with the alignment seismic identified to the north and east of Bolsena Lake \cite{Buonasorte:2013}. During the course of volcanic Pleistocene, therefore, transverse structures trending anti-Apennines were activated, with a major role in guiding the ascent of magmas potash. Next figure shows were major geological structures of the area in question (sedimentary basins, volcanic-tectonic structures and structural highs supposed/known) on the basis of the information gravimetric; many of these show evidence on the surface or seismic. \\ \\ \includegraphics[width=0.5\textwidth{}]{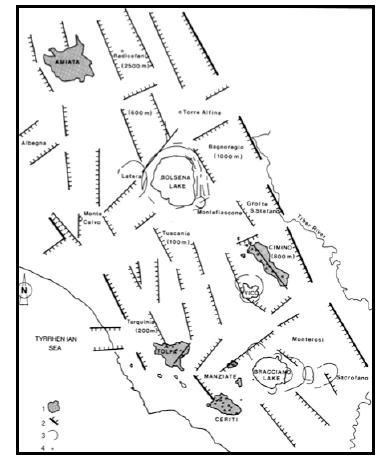} \\
{\scriptsize \bf \textit{Diagram of the main sedimentary basins of Pliocene and Pleistocene, from literature data and the interpretation of the gravity data. 1) Complex volcanic acids Pliocene and Pleistocene. 2) Facilities bordiere major sedimentary basins of Pliocene and Pleistocene, deduced by gravimetric discontinuity. 3) Main facilities and caldera crater. 4) Main lava domes.}} 
\subsection{\normalsize Earthquake}
\normalsize This study evidence an increase of Earthquake activity for 2006 to 2014. \\ In next figure we can see the number of Earthquake in Acquapendente, and table evidence the number of Earthquake event by year. \\ \\
\includegraphics[width=0.5\textwidth{}]{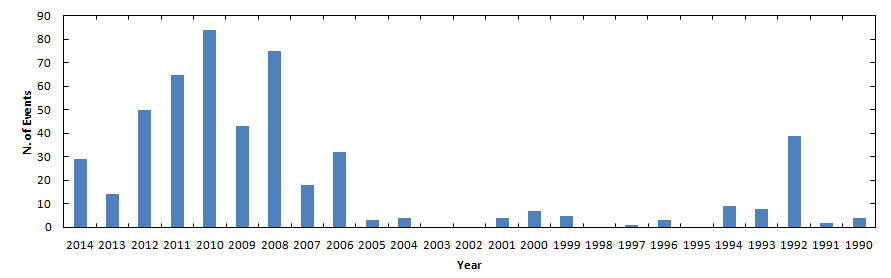} \\
{\scriptsize \bf \textit{Number of Earthquake for 1990 to 2014}} \\
\begin{tabular}{|r|c|c|c|}
\hline
\scriptsize \textbf{Year} &\scriptsize \textbf{N. of Events} &\scriptsize \textbf{ Year} & \scriptsize \textbf{N. of Events}\\
\hline
\scriptsize 2014 & \scriptsize 29 & \scriptsize 2013 & \scriptsize14 \cr
\scriptsize 2012 & \scriptsize 50 & \scriptsize 2011 & \scriptsize 65 \cr
\scriptsize 2010 & \scriptsize 84 & \scriptsize 2009 & \scriptsize 43 \cr
\scriptsize 2008 & \scriptsize 75 & \scriptsize 2007 &\scriptsize 18 \cr
\scriptsize 2006 & \scriptsize 32 &\scriptsize 2005 & \scriptsize 3 \cr
\scriptsize 2004 & \scriptsize 4 & \scriptsize 2003 & \scriptsize 0 \cr
\scriptsize 2002 & \scriptsize 0 & \scriptsize 2001 & \scriptsize 4 \cr
\scriptsize 2000 &\scriptsize 7 & \scriptsize 1999 & \scriptsize 5 \cr
\scriptsize 1998 & \scriptsize 0 & \scriptsize 1997 & \scriptsize 1 \cr
\scriptsize 1996 & \scriptsize 3 &\scriptsize 1995 & \scriptsize 0 \cr
\scriptsize 1994 & \scriptsize 9 & \scriptsize 1993 & \scriptsize 8 \cr
\scriptsize 1992 & \scriptsize 39 &\scriptsize 1991 & \scriptsize 2 \cr 
\scriptsize 1990 &\scriptsize 4 & & \\
\hline
\end{tabular} \\ \\
\scriptsize {\bf {Number of Earthquake in Acquapendente by Year}} \\ \\
\normalsize The study evidence that the major number of Earthquake are production on deep from 5 to 10 Km. Next graphics evidence this. \\ \\
\includegraphics[width=0.5\textwidth{}]{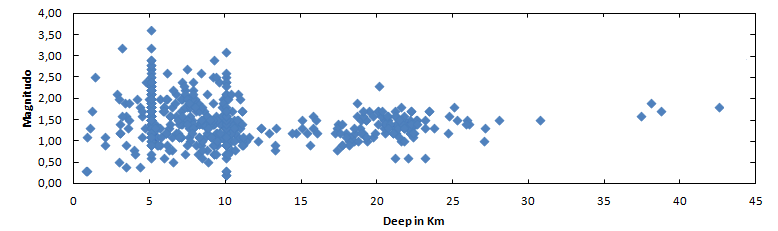} \\
{\scriptsize \bf \textit{Magnitudo/Deep for 1990 to 2014}}
\subsection{\normalsize Wind speed} 
\normalsize In this section we describe the wind speed. \\ In the above image you can see the map of wind speed insistent on the territory of Acquapendente, as seen from the map generated from Atlas wind \cite{atlantevento:2013}. \\ The study of daily wind speed has allowed to estimate on a monthly basis throughout the period included in this study:
\begin{itemize}
\item a decrease in the average for the period 1990-2014 in the speed of the winds, with values ranging from 9 Km/h (in September), and 21 Km/h in the month of July;
\item an increase with higher gusts, in high winds, especially in the period between the autumn and winter, with values between 25 km/h (in August) and 36 Km/h (November).
\end{itemize} 
\normalsize The study also shows that the month of September is the one with the trend towards greater variation in the wind speed, as well it appears from the comparison chart, maximum and minimum wind speed for this study shown in Figure 12. \\ \\ 
\includegraphics[width=0.4\textwidth{}]{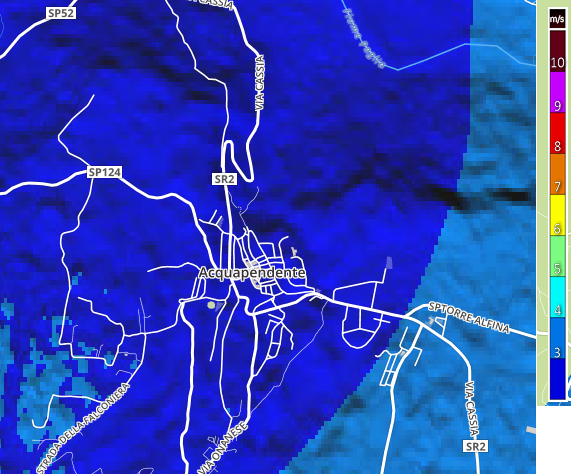} \\ 
{\scriptsize \bf \textit{Wind Speed Atlant of Acquapendente \cite{atlantevento:2013}}}  
\section{\normalsize Conclusion} 
We presented for the first time an analysis of geological and longterm temperature data directly obtained from Acquapendente local meteorological site, inside urban concentration and well above the inversion layer. \\
The data analysis above all those related to annual rainfall, to their regimen and the change of these indices over time indicates that the area is recorded increase the contribution of rainfall and its concentration in the most rainy periods. \\
The extremes of this trend over time if confirmed, would lead to a radical change in climate of the area that would no longer be characterized by the absence of the dry season, but rather by a rainy season with heavy and often heavy rainfall and a dry season with characteristics similar to those of the dry seasons of the Mediterranean climate. \\ The inversion of Acquapendente arouses contrasting effects among citizens, as it is a phenomenon particularly popular in the middle of a hot summer when at least in the early morning you are able to get much lower temperatures and refreshing, while in winter it is seen as a phenomenon inconvenient dates the low temperatures and possible fog and frost that usually occur in parallel. \\ Of course it is undoubtedly a characteristic phenomenon, which has advantages such as the cool summer morning or the ability to create a favorable condition to the snow fall. \\
Given the geology and structure of rocks and outcrops in the area, plus the non-poor seismic activity in the area, went increasing from 2006 until the preparation of this study, as shown in the table in section 4.2 of this paper, it is recommended further deepen the geological knowledge of the area in order to prepare the necessary protections for the populations and infrastructure present.
\section{\normalsize Acknowledgments}
We would like to thank Silvia Gargano for her helpful suggestions and support to made the paper more complete. The constructive comments are highly appreciated. 
\end{multicols}
\newpage
\begin{figure}
\begin{center}
\includegraphics[width=0.8\textwidth]{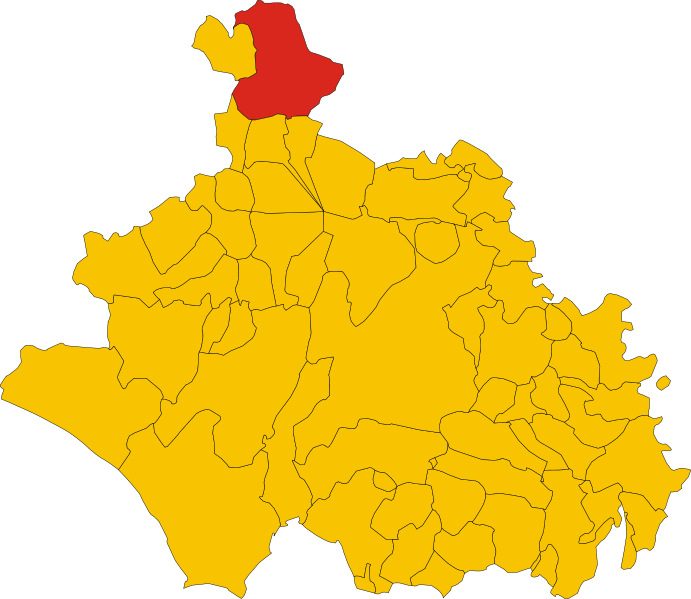}
\caption{Identification of Acquapendente in the Viterbo's  province}
\end{center}
\end{figure}
\centering
\newpage
\begin{table}
\tiny
\centering
\caption{Comparison of temperature on decadal scale} 
\begin{tabular}{|c|c|c|c|c|c|c|c|c|}
\hline
\bf Month & \bf Year &\bf Temperatures Min - Max&\bf Month & \bf Year &\bf  Temperatures Min - Max &\bf Month & \bf Year &\bf Temperatures Min - Max \cr

\hline
\bf Jannary &2014&5 - 11& \bf Febbrary &2014&6 - 13 &\bf March &2014&5 - 16\cr
& 2013 &3 - 10 & &2013 &1 -8 &&2013 &5 - 12 \cr 
& 2012&2 - 11 && 2012&0 - 4 & & 2012&6 - 19 \cr
& 2011&3 - 10 && 2011&3 - 12 & & 2011&5 - 13 \cr
& 2010&2 - 8 && 2010&4 - 10 & & 2010&4 - 13 \cr
& 2009&4 - 9 && 2009&3 - 10 & & 2009&6 - 14 \cr
& 2008&4 - 12 && 2008&3 - 12 & & 2008&5 - 14 \cr
& 2007&4 - 13 && 2007&5 - 14 & & 2007&6 - 16 \cr
& 2006&0 - 8 && 2006&2 - 11 & & 2006&4 - 13 \cr
& 2005&1 - 9 && 2005&0 - 8 & & 2005&3 - 14 \cr
& 2004&1 - 9 && 2004&3 - 12 & & 2004&4 - 13 \cr
& 2003&3 - 10 && 2003&-1 - 9 & & 2003&3 - 16 \cr
& 2002&0 - 10 && 2002&5 - 14 & & 2002&6 - 18 \cr 
& 2001&6 - 11 && 2001&3 - 11 && 2001&8 -16  \cr 
& 2000&1 - 9 && 2000&4 - 13 && 2000&6 - 14  \cr
& 1999&1 - 10 && 1999&1 - 10 && 1999&5 - 14  \cr
& 1998&4 - 11 && 1998&4 - 14 && 1998&4 - 13  \cr
& 1997&4 - 11 && 1997&3 - 12 && 1997&4 - 15  \cr
& 1996&3 - 10 && 1996&0 - 8 && 1996&2 - 10  \cr
& 1995&1 - 9 && 1995&4 - 12 && 1995&3 - 12  \cr
& 1994&4 - 10 && 1994&3 - 11 && 1994&4 - 16  \cr
& 1993&2 - 10 && 1993&-1 - 11 && 1993&2 - 12  \cr
& 1992&2 - 10 && 1992&2 - 12 && 1992&4 - 14  \cr
& 1991&1 - 10 && 1991&1 - 9 && 1991&6 - 16  \cr
& 1990&1 - 10 && 1990&3 - 13 && 1990&4 - 14  \cr\hline
\bf April &2014&9 - 18& \bf May &2014&11 - 21&\bf June &2014&16 - 27\cr
& 2013&9 - 18 && 2013 &10 - 19 && 2013 &15 - 26 \cr 
& 2012&8 - 17 & & 2012&11 - 21 & & 2012&17 - 30 \cr
& 2011&9 - 19 & & 2011&12 - 24 & & 2011&16 - 27 \cr
& 2010&7 - 18 & & 2010&12 - 20& & 2010&15 - 25 \cr
& 2009&10 - 18 & & 2009&15 - 25 & & 2009&18 - 26 \cr
& 2008&7 - 17 & & 2008&11 - 22 & & 2008&18 - 26 \cr
& 2007&9 - 22 & & 2007&12 - 24 & & 2007&16 - 24 \cr
& 2006&8 - 19 & & 2006&11 - 24 & & 2006&13 - 27 \cr
& 2005&6 - 17 & & 2005&12 - 25 & & 2005&15 - 28 \cr
& 2004&7 - 17 & & 2004&9 - 20 & & 2004&14 - 26 \cr
& 2003&6 - 18 & & 2003&9 - 21 & & 2003&18 - 33 \cr 
& 2002&7 - 20 & & 2002&9 - 19 & & 2002&13 - 25 \cr 
& 2001&5 - 16 & & 2001&12 - 23 & & 2001&14 - 27 \cr
& 2000&10 - 17 && 2000&14 - 24 && 2000&14 - 29  \cr
& 1999&8 - 16 && 1999&14 - 23 && 1999&18 - 28  \cr
& 1998&8 - 17 && 1998&13 - 22 && 1998&16 - 27  \cr
& 1997&4 - 14 && 1997&11 - 23 && 1997&15 - 26  \cr
& 1996&6 - 16 && 1996&10 - 20 && 1996&14 - 27  \cr
& 1995&4 - 14 && 1995&8 - 20 && 1995&11 - 24  \cr
& 1994&5 - 15 && 1994&10 - 21 && 1994&14 - 25  \cr
& 1993&6 - 17 && 1993&10 - 23 && 1993&14 - 27  \cr
& 1992&7 - 16 && 1992&11 - 23 && 1992&13 - 23  \cr
& 1991&3 - 14 && 1991&6 - 16 && 1991&12 - 24  \cr
& 1990&4 - 13 && 1990&10 - 21 && 1990&14 - 25  \cr\hline
\bf July &2014&7 - 27& \bf August &2014&17 - 28&\bf September &2014&15 - 25\cr
&2013 &19 - 30  &&2013 &19 - 31  &&2013 &15 - 26 \cr 
& 2012&19 - 33 & & 2012&20 - 34 & & 2012&16 - 25 \cr
& 2011&17 - 27 & & 2011&18 - 31 & & 2011&16 - 28 \cr
& 2010&19 - 31 & & 2010&17 - 29 & & 2010&14 - 24 \cr
& 2009&20 - 30 & & 2009&20 - 32 & & 2009&16 - 26 \cr
& 2008&19 - 30 & & 2008&19 - 32 & & 2008&15 - 25 \cr
& 2007&17 - 32 & & 2007&17 - 30 & & 2007&14 - 26 \cr
& 2006&18 - 33 & & 2006&17 - 28 & & 2006&16 - 27 \cr
& 2005&17 - 31 & & 2005&16 - 28 & & 2005&14 - 25 \cr
& 2004&16 - 30 & & 2004&17 - 30 & & 2004&14 - 27 \cr
& 2003&18 - 34 & & 2003&19 - 33 & & 2003&14 - 27 \cr 
& 2002&16 - 29 & & 2003&16 - 27 & & 2003&13 - 23 \cr 
& 2001&17 - 31 & & 2001&18 - 32 & & 2001&12 - 23 \cr
& 2000&16 - 30 & & 2000&17 - 33 && 2000&14 - 26  \cr
& 1999&20 - 29 && 1999&20 - 31 && 1999&16 - 26  \cr
& 1998&19 - 31 && 1998&19 - 31 && 1998&14 - 24  \cr
& 1997&15 - 29 && 1997&17 - 29 && 1997&15 - 27  \cr
& 1996&16 - 29 && 1996&16 - 28 && 1996&11 - 21  \cr
& 1995&17 - 30 && 1995&15 - 26 && 1995&11 - 21  \cr
& 1994&18 - 31 && 1994&19 - 32 && 1994&15 - 25  \cr
& 1993&15 - 28 && 1993&18 - 32 && 1993&14 - 24  \cr
& 1992&16 - 29 && 1992&18 - 32 && 1992&16 - 26  \cr
& 1991&16 - 30 && 1991&17 - 31 && 1991&15 - 25  \cr
& 1990&15 - 29 && 1990&16 - 28 && 1990&13 - 24  \cr\hline
\bf October &2014&13 - 22& \bf November &2014&10 - 17&\bf December &2014&6 - 12\cr
&2013 &13 - 22  &&2013 &8 -14  &&2013 &4 - 11 \cr 
& 2012&12 - 22 & & 2012&9 - 15 & & 2012&2 - 9 \cr
& 2011&11 - 21 & & 2011&7 - 16 & & 2011&5 - 13 \cr
& 2010&10 - 19 & & 2010&7 - 14 & & 2010&6 - 4 \cr
& 2009&10 - 19 & & 2009&7 - 16 & & 2009&5 - 10 \cr
& 2008&12 - 22 & & 2008&8 - 15 & & 2008&4 - 9 \cr
& 2007&11 - 21 & & 2007&5 - 14 & & 2007&2 - 10 \cr
& 2006&12 - 23 & & 2006&7 - 17 & & 2006&5 - 13 \cr
& 2005&10 - 20 & & 2005&4 - 8 & & 2005&2 -9 \cr
& 2004&13 - 22 & & 2004&7 - 14 & & 2004&5 - 12 \cr
& 2003&10 - 19 & & 2003&8 - 15 & & 2003&3 - 10 \cr
& 2002&10 - 20 & & 2003&10 - 17 & & 2003&6 - 11 \cr
& 2001&13 - 24 & & 2001&6 - 15 & & 2001&1 - 8 \cr
& 2000&12 - 20 && 2000&8 - 16 && 2000&5 - 13  \cr
& 1999&13 - 21 && 1999&7 - 14 && 1999&5 - 10  \cr
& 1998&11 - 19 && 1998&6 - 12 && 1998&2 - 9  \cr
& 1997&10 - 19 && 1997&8 - 14 && 1997&5 - 11  \cr
& 1996&9 - 18 && 1996&7 - 14 && 1996&3 - 8  \cr
& 1995&9 - 20 && 1995&4 - 12 && 1995&5 - 10  \cr
& 1994&10 - 19 && 1994&8 - 15 && 1994&4 - 11  \cr
& 1993&11 - 18 && 1993&5 - 11 && 1993&3 - 11  \cr
& 1992&12 - 19 && 1992&7 - 15 && 1992&3 - 10  \cr
& 1991&9 - 17 && 1991&5 - 12 && 1991&-1 - 8  \cr
& 1990&11 - 19 && 1990&5 - 13 && 1990&1 - 7  \cr 
\hline
\end{tabular}
\end{table}
\centering
\begin{table}
\centering \tiny
\caption{Day Time and Night Time Variation} 

\begin{tabular}{|c|c|c|c|c|c|c|c|c|c|c|c|c|}
\hline
\bf Year & \bf Jan &\bf Feb &\bf Mar & \bf Apr &\bf May&\bf Jun & \bf Jul &\bf Aug&\bf Sep&\bf Oct &\bf Nov&\bf Dec \cr
\hline
\bf 2014 & 6 & 7 & 9 & 9 & 10 & 11 & 10 & 11 & 9 & 10 & 8 & 6 \cr  \hline
\bf 2013 & 7 & 7 & 7 & 9 & 9 & 11 & 12 & 12 & 11& 9& 6 & 8 \cr  \hline
\bf 2012 & 9 & 9  & 12  & 9  & 10 & 13  & 13 & 14 & 9 & 9 & 7 & 7 \cr  \hline
\bf 2011 &7 &9  &9  &10  &12 &11  &11 &13 &12 &10 &10 &7 \cr  \hline
\bf 2010 &6 &6  &9  &10  &8 &10  &12 &12 &10 &9 &7 &6 \cr  \hline
\bf 2009 &6 &8  &9  &8  &10 &8  &10 &11 &10 &9 &8 &5 \cr  \hline
\bf 2008 &8 &10 &9  &10  &10 &8  &10 &13 &10 &9 &8 &5 \cr  \hline
\bf 2007 &9 &10  &10  &13  &10 &12  &15 &13 &12 &10 &9 &8 \cr  \hline
\bf 2006 &9 &9  &9  &11  &13 &11  &14 &11 &11 &10 &10 &8 \cr  \hline
\bf 2005 &8 &9  &11  &11  &13 &11  &15 &12 &11 &9 &10 &8 \cr  \hline
\bf 2004 &8 &8  &10  &10  &12 &14  &15 &14 &12 &9 &7 &6 \cr  \hline
\bf 2003 &7 &10  &13  &13  &14 &11  &16 &15 &12 &9 &7 &7 \cr  \hline
\bf 2002 & 10 & 9 & 12 & 13 & 12 & 14 & 13 & 11 & 10 & 10 & 7 & 6 \cr  \hline
\bf 2001 &6  &9  &8  &11  &11  &13  &14  &14  &11  &11  &9  &8  \cr  \hline
\bf 2000 &8  &8  &8  &8  &10  &15  &14  &15  &12  &9  &7  &7  \cr  \hline
\bf 1999 &8  &8  &9  &9  &9  &10  &10  &12  &9  &8  &7  &6  \cr  \hline
\bf 1998 &7  &10  &9  &9  &9  &11  &11  &12  &10  &8  &6  &7  \cr  \hline
\bf 1997 &7  &9  &11  &10  &12  &11  &13  &12  &12  &9  &6  &6  \cr  \hline
\bf 1996 &6  &8 &8  &10  &10  &12  &13  &12  &10  &8  &7  &5  \cr  \hline
\bf 1995 &8  &8  &10  &9  &11  &13  &13  &11  &10  &11  &8  &6  \cr  \hline
\bf 1994 &7  &8  &12  &9  &11  &11  &13  &14  &11  &9  &6  &7  \cr  \hline
\bf 1993 &6  &5  &7  &11  &16  &21  &21  &24  &19  &14  &7  &7  \cr  \hline
\bf 1992 &9  &10  &10  &9  &12  &10  &12  &14  &13  &7  &8  &7  \cr  \hline
\bf 1991 &9  &8  &9  &10  &10  &12  &14  &14  &11  &9  &8  &9  \cr  \hline
\bf 1990 &9  &10  &10  &9  &11  &13  &14  &13  &11  &8  &8  &6  \cr  \hline

\hline
\bf Average 1990-2014&7,47 &8,48  &9,50 &9,96 &11,05 &12,19 &13,10 &12,97 &11,11 &9,30 &7,72 &6,60 \cr  \hline

\end{tabular}
\end{table}


\begin{table}
\tiny
\centering
\caption{Rain/Year [mm]}

\begin{tabular}{|c|c|c|c|c|c|c|c|c|c|c|c|c|c|c|c|}
\hline
\bf Year & & Jan & Feb & Mar & Apr & May & Jun & Jul & Ago & Sep & Oct & Nov &  Dec & \bf Tot & \bf \%Year \\
\hline
&  &  &  & &  & & & & & & & & & &          \cr
\hline
2014 & \bf 82 & 95,00 & 83,00 & 95,00 & 95,00 & 60,00 & 85,00 & 90,00 & 55,00 & 85,00 & 80,00 & 180,00 & 235,00 & \bf 1238,00 & \bf 5,78\% \cr \hline
2013 & \bf 128 & 185,00 & 155,00 & 140,00 & 35,00 & 80,00 & 155,00 & 255,00 &  15,00 & 50,00 & 65,00 & 95,00 & 25,00 & \bf 1255,00 & \bf 5,86\% \cr \hline
2012 & \bf 58 &  80,43 &  75,00 & 55,00 & 120,00 &  80,00 & 15,00 & 5,00 &  35,00 & 125,00 &  115,00 &  95,00 & 145,00 & \bf 945,43 & \bf 4,42\% \cr
\hline
2011 & \bf 86 &  40,00 &  35,00 &  125,00 & 40,00 & 144,00 & 140,00 &  150,00 & 15,00 &  35,00 &  45,00 & 35,00 & 55,00 & \bf 859,00 & \bf 4,01\% \cr
\hline
2010 &\bf  49 &  48,00 &  67,00 & 35,00 &  67,00 & 96,00 & 44,00 &  10,00 &  25,00 & 5,00 & 47,00 &  107,00 & 53,00 & \bf 604,00 & \bf 2,82\%  \cr
\hline
2009 &   \bf 96 & 44,30 & 95,10 & 105,40 & 76,50 & 60,60 & 154,50 & 92,30 & 140,00 & 264,00 & 190,00 & 146,00 & 45,00 & \bf 1413,70 & \bf 6,61\% \cr
\hline
2008 &  \bf 98 & 71,00 & 53,50 & 93,00 & 104,00 & 146,20 & 81,90 & 32,80 & 99,70 & 51,40 & 146,40 & 285,90 & 16,00 & \bf 1181,80 & \bf 5,52\% \cr
\hline
2007 &   \bf 56 & 47,40 & 74,50 & 120,90 & 50,10 & 47,20 & 60,00 & 57,60 & 58,50 & 48,50 & 68,50 & 24,70 & 18,90 & \bf 676,80 & \bf 3,16\% \cr
\hline
2006 &   \bf 82 &232,20 & 124,80 & 94,00 & 47,80 & 19,90 & 10,90 & 110,10 & 123,20 & 98,10 & 43,20 & 77,20 & 3,00 & \bf 985,30 & \bf 4,60\% \cr
\hline
2005 &   \bf 76 & 44,40 & 38,20 & 51,70 & 91,70 & 93,40 & 54,80 & 24,60 & 130,70 & 202,60 & 101,30 & 59,30 & 17,20 & \bf 909,90 & \bf 4,25\% \cr
\hline
2004 &   \bf 84 &  41,70 & 123,50 & 97,00 & 210,40 & 50,60 & 1,50 &  23,10 &  35,00 & 63,00 & 101,60 & 212,00 & 53,60 & \bf 1013,00 & \bf 4,73\% \cr
\hline
2003 & \bf 63 & 84,20 & 24,40 & 17,00 & 23,20 & 140,10 & 57,10 & 53,90 & 25,70 & 95,40 & 94,40 & 115,80 & 20,80 &\bf 752,00 & \bf 3,51\% \cr
\hline
2002 & \bf 64 & 32,00 & 37,70 & 11,40 & 43,50 & 63,20 & 34,10 & 76,20 & 110,80 & 132,30 & 18,20 & 82,70 & 123,20 &\bf 765,30 & \bf 3,58\% \cr
\hline
2001 & \bf43  &142,20  &34,90  &83,30  &74,20  &27,30  &17,00  & 20,60 &8,00  &22,50  &11,60  &62,60  &13,80  &\bf518,00  & \bf 2,42 \% \cr
\hline
2000 & \bf 57 &15,00  &7,00  &38,00  &30,00  &98,00  &37,60  &59,10  &57,20  &26,00  &125,40  &108,80  &84,60  &\bf 686,70 & \bf 3,21\% \cr
\hline
1999 & \bf 58  &20,00  &15,00  &70,00  &60,00  &55,00  &85,00  & 85,00 &35,00  &100,00  &45,00  &75,00  &50,00  &\bf 695,00 & \bf 3,25\% \cr
\hline
1998 & \bf 47 &30,00  &20,00  &20,00  &100,00  &35,00  &60,00  & 5,00 &60,00  &55,00  &90,00  &45,00  &45,00  &\bf 565,00 & \bf 2,64\% \cr
\hline
1997 & \bf 41 &50,00  &30,00  &25,00  &65,00  &10,00  &50,00  & 20,00 &45,00  &10,00  &90,00  &55,00 & 45,00 &\bf 495,00 & \bf 2,31\% \cr
\hline
1996 & \bf 88 &65,00  &115,00  &30,00  &115,00  &145,00  &80,00  &30,00  &65,00  &147,00  &58,00  &132,00  &70,00  &\bf 1052,00 & \bf 4,92\% \cr
\hline
1995 & \bf 84 &45,00  &60,00  &70,00  &93,00  &105,00  &20,00  & 30,00 &280,00  &90,00  &15,00  &30,00  &170,00  &\bf 1008,00 & \bf 4,71\% \cr
\hline
1994 & \bf 50 &37,00  &30,00  &5,00  &163,00  &20,00  &85,00  & 45,00 &10,00  &62,00  &50,00  &40,00  &50,00  &\bf 597,00 & \bf 2,79\% \cr
\hline
1993 & \bf 50 &10,00  &5,00  &25,00  &75,00  &160,00  &15,00  & - &20,00  &100,00  &70,00  &80,00  &45,00  &\bf 605,00 & \bf 2,83\% \cr
\hline
1992 & \bf 78 &50,00  &15,00  &85,00  &100,00  &65,00  &115,00  &30,00  &30,00  &65,00  &242,00  &50,00  &85,00  &\bf 932,00 & \bf 4,35\% \cr
\hline
1991 & \bf 69 &10,00  &55,00  &45,00  &110,00  &110,00  &30,00  &60,00  &20,00  &110,00  &140,00  &128,00  &5,00  &\bf 823,00 & \bf 3,85\% \cr
\hline
1990 & \bf 69 &15,00  &16,00  &25,00  &140,00  &90,00  &10,00  & 45,00 &85,00  &35,00  &175,00  &115,00  &75,00  &\bf 826,00 & \bf 3,86\% \cr
\hline
\bf Average 1990-2014 & \bf 70,25 & \bf 61,39 & \bf 55,58 & \bf 62,47 & \bf 85,21 & \bf 80,06 & \bf 59,94 & \bf 56,41 & \bf 63,35 & \bf 83,11 & \bf 89,10 & \bf 97,48 & \bf 61,96 & \bf 21401,93 &  \cr
\hline
\bf &&&&&&&&&&&&&& \bf Year Average & \bf 856,08 \\
\hline
\end{tabular}
\end{table}
\begin{table}
\caption{Wind Average[1990-2014]/Year on Km/h}
\tiny
\begin{tabular}{|c|c|c|c|c|c|c|c|c|c|c|c|c|c|c|c|} 
\hline
\multicolumn{ 14}{|c}{} &            &            \\
\hline
\multicolumn{ 1}{|c|}{Day} &        Jan &        Feb &        Mar &        Apr &        May &        Jun &        Jul &        Ago &        Sep &        Oct &        Nov &        Dec &\bf Average &  \bf Min &  \bf Max  \\
\hline
\multicolumn{ 1}{|c|}{} &            &            &            &            &            &            &            &            &            &            &            &            &            &            &            \\

         1 &      29,09 &      26,35 &      24,45 &      26,38 &      20,10 &      23,64 &      21,59 &      23,43 &      26,91 &      18,43 &      23,17 &      27,09 & {\bf 24,22} &      18,43 &      29,09  \cr
         2 &      23,18 &      24,57 &      27,23 &      24,33 &      21,33 &      21,24 &      23,14 &      23,33 &      23,96 &      18,83 &      24,70 &      22,61 & {\bf 23,20} &      18,83 &      27,23  \cr 
         3 &      26,32 &      22,13 &      26,50 &      22,54 &      25,86 &      22,45 &      22,61 &      23,61 &      22,91 &      20,00 &      22,57 &      24,82 & {\bf 23,53} &      20,00 &      26,50  \cr
         4 &      26,95 &      24,74 &      27,64 &      26,79 &      25,29 &      21,77 &      23,04 &      23,91 &      24,46 &      22,39 &      24,17 &      22,18 & {\bf 24,45} &      21,77 &      27,64  \cr 
         5 &      24,09 &      25,26 &      29,10 &      27,04 &      19,23 &      21,82 &      22,86 &      25,09 &      23,87 &      22,74 &      22,91 &      20,82 & {\bf 23,73} &      19,23 &      29,10  \cr
         6 &      21,77 &      26,74 &      24,90 &      22,83 &      23,95 &      22,57 &      22,52 &      23,71 &      22,25 &      20,96 &      21,77 &      24,45 & {\bf 23,20} &      20,96 &      26,74 \cr 
         7 &      19,27 &      24,00 &      27,41 &      24,63 &      23,24 &      21,95 &      24,65 &      22,26 &      24,00 &      24,29 &      26,65 &      28,09 & {\bf 24,20} &      19,27 &      28,09  \cr 
         8 &      19,86 &      26,43 &      29,38 &      22,00 &      20,50 &      22,14 &      25,26 &      23,86 &      22,82 &      22,21 &      23,57 &      26,18 & {\bf 23,69} &      19,86 &      29,38  \cr  
         9 &      18,55 &      25,61 &      25,71 &      25,22 &      20,00 &      22,81 &      22,83 &      22,29 &      26,13 &      19,88 &      21,22 &      28,71 & {\bf 23,25} &      18,55 &      28,71  \cr  
        10 &      18,59 &      24,91 &      26,14 &      26,30 &      19,65 &      22,95 &      23,52 &      24,55 &      22,83 &      20,71 &      23,65 &      26,68 & {\bf 23,37} &      18,59 &      26,68  \cr 
        11 &      21,13 &      25,86 &      26,76 &      27,09 &      18,43 &      23,00 &      24,83 &      24,52 &      22,95 &      19,67 &      25,38 &      20,86 & {\bf 23,37} &      18,43 &      27,09  \cr 
        12 &      23,26 &      24,57 &      22,00 &      26,70 &      21,60 &      21,24 &      23,52 &      25,75 &      28,62 &      21,17 &      22,48 &      21,50 & {\bf 23,53} &      21,17 &      28,62  \cr 
        13 &      21,68 &      23,35 &      23,64 &      24,22 &      21,10 &      21,52 &      22,43 &      22,29 &      24,50 &      21,91 &      23,41 &      23,24 & {\bf 22,77} &      21,10 &      24,50  \cr 
        14 &      22,30 &      21,39 &      20,41 &      23,91 &      22,50 &      21,76 &      22,82 &      24,04 &      26,50 &      22,39 &      25,50 &      26,10 & {\bf 23,30} &      20,41 &      26,50  \cr 
        15 &      25,09 &      27,59 &      21,55 &      24,04 &      22,71 &      23,67 &      25,27 &      22,88 &      23,26 &      23,22 &      26,27 &      24,86 & {\bf 24,20} &      21,55 &      27,59  \cr 

        16 &      23,17 &      28,77 &      24,68 &      25,48 &      21,70 &      22,71 &      24,77 &      22,24 &      23,48 &      22,43 &      31,77 &      23,77 & {\bf 24,58} &      21,70 &      31,77  \cr 
        17 &      23,04 &      25,96 &      23,82 &      23,70 &      23,05 &      22,64 &      24,39 &      22,54 &      21,00 &      22,74 &      26,27 &      23,91 & {\bf 23,59} &      21,00 &      26,27  \cr 
        18 &      21,52 &      25,82 &      24,14 &      24,61 &      26,25 &      21,52 &      23,22 &      20,71 &      21,17 &      20,83 &      23,77 &      22,00 & {\bf 22,96} &      20,71 &      26,25  \cr 
        19 &      22,82 &      25,87 &      25,86 &      23,52 &      26,05 &      22,00 &      23,21 &      22,25 &      20,32 &      22,39 &      23,64 &      21,77 & {\bf 23,31} &      20,32 &      26,05  \cr 
        20 &      21,86 &      23,61 &      22,18 &      21,30 &      22,95 &      23,38 &      24,30 &      22,17 &      22,54 &      22,52 &      25,36 &      24,27 & {\bf 23,04} &      21,30 &      25,36  \cr 
        21 &      24,36 &      22,39 &      23,77 &      24,95 &      21,65 &      21,23 &      25,30 &      23,50 &      24,17 &      26,96 &      23,91 &      23,35 & {\bf 23,80} &      21,23 &      26,96  \cr 
        22 &      25,10 &      24,36 &      24,64 &      21,00 &      21,15 &      26,36 &      23,30 &      22,33 &      20,52 &      21,70 &      22,00 &      23,83 & {\bf 23,02} &      20,52 &      26,36  \cr 
        23 &      25,22 &      26,27 &      26,78 &      21,95 &      23,76 &      22,48 &      23,48 &      23,09 &      23,04 &      24,09 &      21,59 &      27,39 & {\bf 24,10} &      21,59 &      27,39  \cr 
        24 &      28,74 &      24,05 &      25,43 &      21,17 &      22,80 &      21,48 &      27,30 &      21,26 &      23,71 &      22,57 &      21,95 &      28,45 & {\bf 24,08} &      21,17 &      28,74  \cr 
        25 &      29,48 &      23,00 &      27,00 &      22,61 &      22,62 &      22,48 &      23,17 &      22,48 &      23,00 &      20,65 &      20,14 &      27,32 & {\bf 23,66} &      20,14 &      29,48  \cr 
        26 &      26,70 &      23,27 &      26,48 &      22,96 &      21,60 &      22,17 &      22,74 &      21,52 &      25,57 &      22,29 &      21,68 &      28,65 & {\bf 23,80} &      21,52 &      28,65  \cr 
        27 &      23,43 &      23,45 &      28,65 &      21,78 &      22,00 &      22,70 &      23,09 &      22,54 &      23,04 &      20,26 &      23,24 &      24,52 & {\bf 23,23} &      20,26 &      28,65  \cr 
        28 &      21,22 &      27,59 &      26,13 &      18,41 &      21,45 &      23,65 &      23,52 &      24,86 &      21,91 &      21,78 &      22,77 &      26,35 & {\bf 23,30} &      18,41 &      27,59  \cr 
        29 &      26,48 &      20,20 &      23,78 &      20,86 &      24,14 &      23,04 &      23,61 &      24,22 &      22,22 &      22,55 &      21,73 &      25,00 & {\bf 23,15} &      20,20 &      26,48  \cr 
        30 &      26,04 &            &      25,08 &      20,95 &      21,62 &      21,96 &      23,35 &      24,77 &      21,95 &      22,74 &      22,82 &      23,36 & {\bf 23,15} &      20,95 &      26,04  \cr 
        31 &      25,52 &            &      28,33 &      18,50 &      21,24 &      20,00 &      25,00 &      25,83 &      12,00 &      23,77 &      27,25 &      26,64 & {\bf 23,10} &      12,00 &      28,33  \cr 
\hline
{\bf Average} & {\bf 23,74} & {\bf 24,76} & {\bf 25,47} & {\bf 23,48} & {\bf 22,24} & {\bf 22,40} & {\bf 23,70} & {\bf 23,29} & {\bf 23,08} & {\bf 21,90} & {\bf 23,78} & {\bf 24,80} & {\bf Average} & {\bf 12,00} & {\bf 31,77} \\
\hline 
 {\bf Min} &      18,55 &      20,20 &      20,41 &      18,41 &      18,43 &      20,00 &      21,59 &      20,71 &      12,00 &      18,43 &      20,14 &      20,82 & {\bf 22,77} &            &            \\
\hline 
 {\bf Max} &      29,48 &      28,77 &      29,38 &      27,09 &      26,25 &      26,36 &      27,30 &      25,83 &      28,62 &      26,96 &      31,77 &      28,71 & {\bf 24,58} &            &            \\
\hline
\end{tabular}  
\end{table}
\centering
\begin{table}
\centering \tiny
\caption{Trend of Temperature of Acquapendente in Jan-Feb-Mar and Jun-Jul-Aug} 
\begin{tabular}{|c|c|c|c|c|c|c|c|c|}
\hline
\bf Year & \bf Jan & \bf Feb &\bf Mar&\bf Jun & \bf Jul &\bf Aug \cr
\hline
\bf 2014 & 7,68  & 9,79  & 10,06  &21,37 &22,15  &22,84 \cr  \hline
\bf 2013 &6,06  &4,66  &8,58  &20,78  &24,50  &25,19 \cr  \hline
\bf 2012 &6,35  &1,42  &12,58  &23,48  &25,90  &27,08 \cr  \hline
\bf 2011 &6,53 &7,23  &8,54  &21,93  &22,11  &23,69 \cr  \hline
\bf 2010 &5,10  &6,79  &8,56  &19,48  &24,90  &23,31 \cr  \hline
\bf 2009 &6,48  &6,21  &10,08  &21,62  &25,03  &25,95 \cr  \hline
\bf 2008 &7,77  &7,51  &9,34  &21,57  &24,61  &25,53 \cr  \hline
\bf 2007 &8,71  &9,59  &10,71  &22,09  &24,74  &24,15 \cr  \hline
\bf 2006 &4,04  &6,44  &8,44  &20,29  &25,58  &22,76 \cr  \hline
\bf 2005 &4,76  &3,99  &8,76  &21,44  &24,26  &21,89 \cr  \hline
\bf 2004 &4,88  &7,48  &8,28  &19,38  &24,74  &23,98 \cr  \hline
\bf 2003 &6,42  &3,94  &9,47  &25,33  &26,27  &28,27 \cr  \hline
\bf 2002 &4,00  &7,65  &12,26  &23,85 &23,15  &21,45 \cr  \hline
\bf 2001 &8,47  &7,13  &11,74  &20,43 &23,96  &25,06 \cr  \hline
\bf 2000 &5,35  &8,40  &10,05  &21,36  &22,87  &24,93 \cr  \hline
\bf 1999 &5,73  &5,51  &9,37  &23,22  &24,48  &25,52 \cr  \hline
\bf 1998 &7,06  &8,67  &8,55  &21,92  &24,95  &25,03 \cr  \hline
\bf 1997 &7,43  &7,57  &9,31  &20,18  &22,13  &23,13 \cr  \hline
\bf 1996 &6,62  &3,91  &5,94  &20,36  &22,26  &22,00 \cr  \hline
\bf 1995 &4,79  &8,13  &7,45  &17,10  &23,82  &20,84 \cr  \hline
\bf 1994 &6,95  &6,75  &10,05  &19,20  &24,42  &25,39 \cr  \hline
\bf 1993 &5,88  &5,41  &7,14  &20,53  &21,60  &24,65 \cr  \hline
\bf 1992 &6,00  &6,81  &9,05  &18,30  &22,50  &25,16 \cr  \hline
\bf 1991 &5,62  &5,13 &10,96  &18,03  &23,13  &23,71  \cr  \hline
\bf 1990 &5,77  &7,82  &9,23  &18,76  &22,10  &21,98  \cr 
\hline
\bf Average 1990-2014 &6,18 &6,56 &9,38 &20,88 &23,85  &24,14  \cr  \hline

\end{tabular}
\end{table}
\newpage \clearpage

\begin{figure}
\begin{center}
\includegraphics[width=0.5\textwidth]{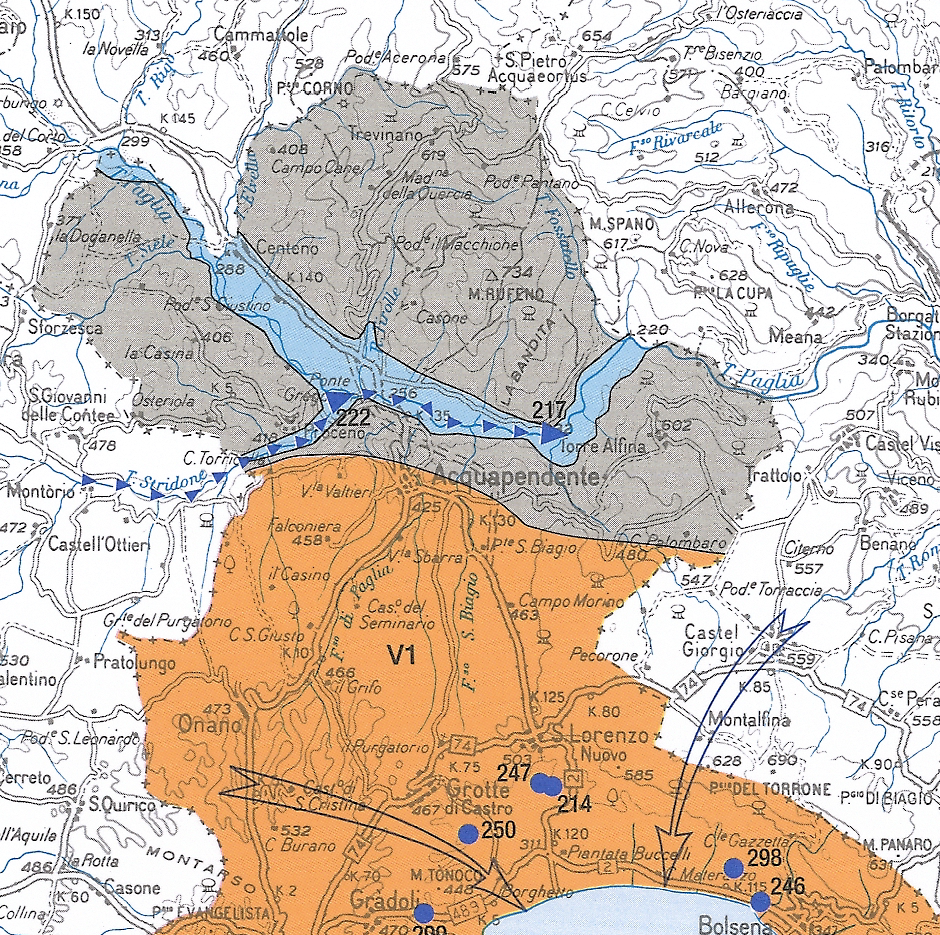} \quad 
\includegraphics[width=0.2\textwidth]{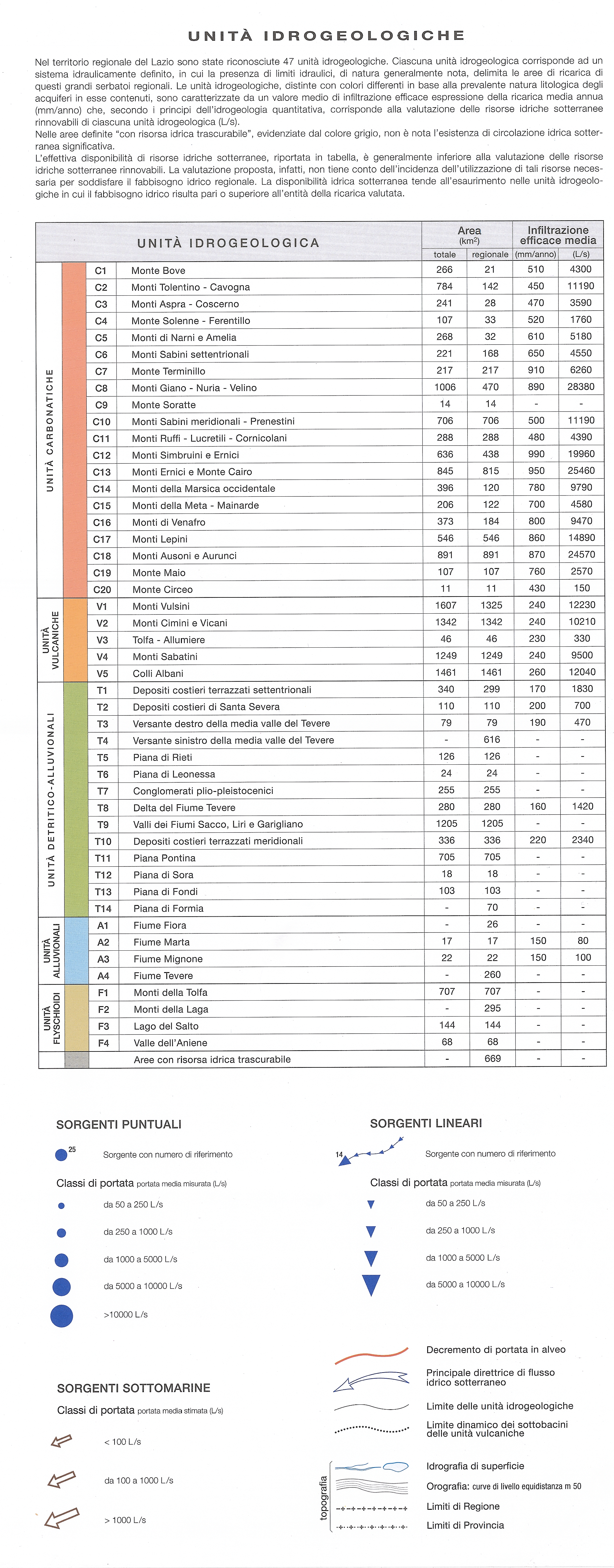}
\caption{Idro-Geological Map of Acquapendente - VT \cite{ISPRA:2013}} 
\end{center}
\end{figure}

\begin{figure}
\begin{center}
\includegraphics[width=1.3\textwidth{}]{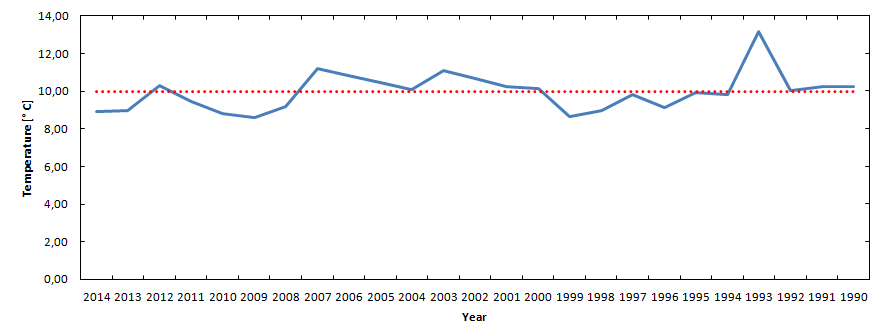}
\caption{Day time and night time variations (solid line) and Average Day time and Night time variations 1990-2014 (dotted line)}
\end{center}
\end{figure}
\begin{figure}
\begin{center}
\includegraphics[width=1.3\textwidth{}]{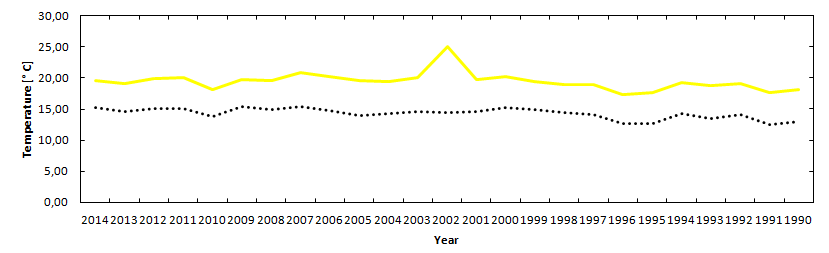}
\caption{Maximum Temperature (solid line) and Average Temperature Variations 1990-2014 (dotted line)}
\end{center}
\end{figure}
\begin{figure}
\begin{center}
\includegraphics[width=1.3\textwidth{}]{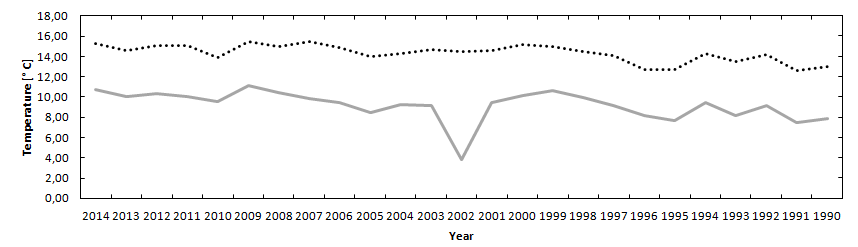}
\caption{Minimum Temperature (solid line) and Average Temperature Variations 1990-2014 (dotted line)}
\end{center}
\end{figure}
\begin{figure}
\begin{center}
\includegraphics[width=1.3\textwidth{}]{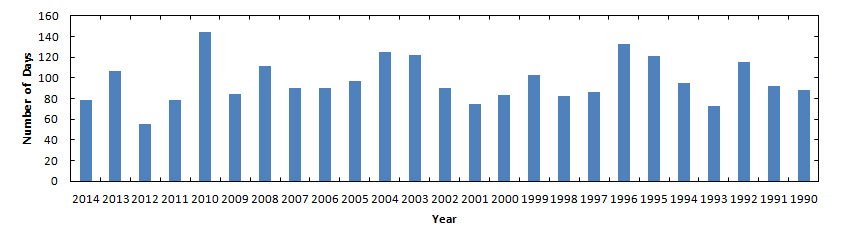}
\caption{Total Days of Rain by Year}
\end{center}
\end{figure}
\begin{figure}
\begin{center}
\includegraphics[width=1.3\textwidth{}]{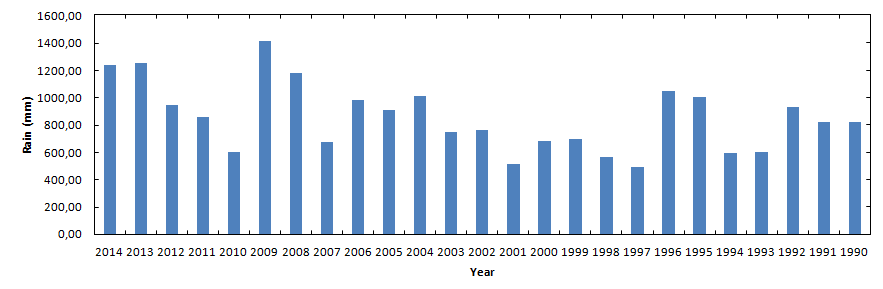}
\caption{Total rain [mm] by year}
\end{center}
\end{figure}
\begin{figure}
\begin{center}
\includegraphics[width=1.3\textwidth{}]{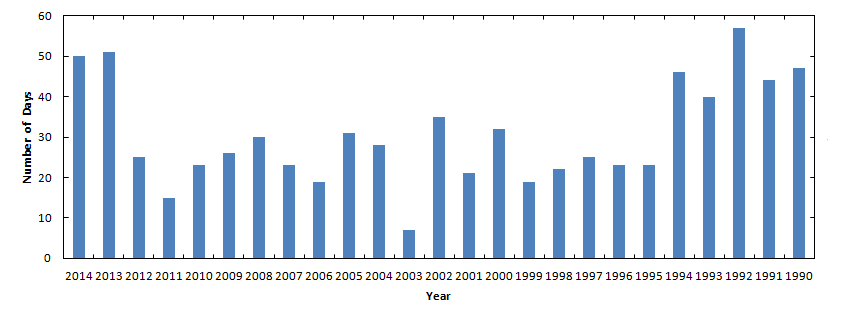}
\caption{Total day of Fog by year}
\end{center}
\end{figure}
\begin{figure}
\begin{center}
\includegraphics[width=1.3\textwidth{}]{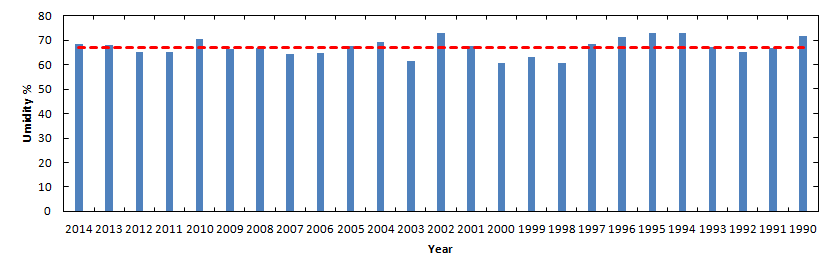}
\caption{Umidity (solid line) and Average Umidity Variations 1990-2014 (dotted line) }
\end{center}
\end{figure}
\begin{figure}
\begin{center}
\includegraphics[width=1.3\textwidth{}]{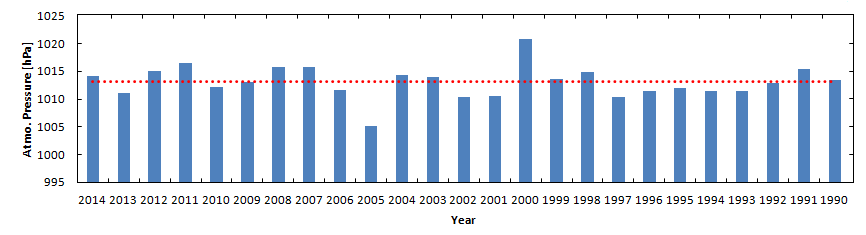}
\caption{Pressure (solid line) and Average Pressure Variations 1990-2014 (dotted line) }
\end{center}
\end{figure}
\begin{figure}
\begin{center}
\includegraphics[width=1.3\textwidth{}]{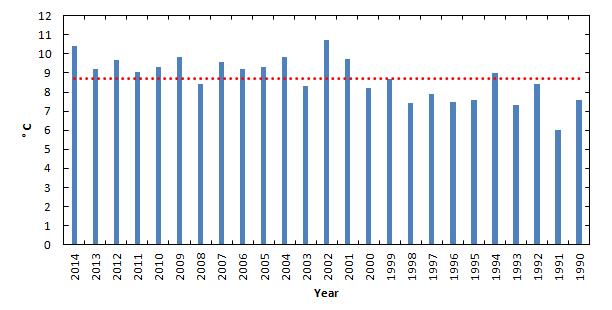}
\caption{Dew Point (solid line) and Average Dew Point Variations 1990-2014 (dotted line) }
\end{center}
\end{figure}
\begin{figure}
\begin{center}
\includegraphics[width=1.3\textwidth{}]{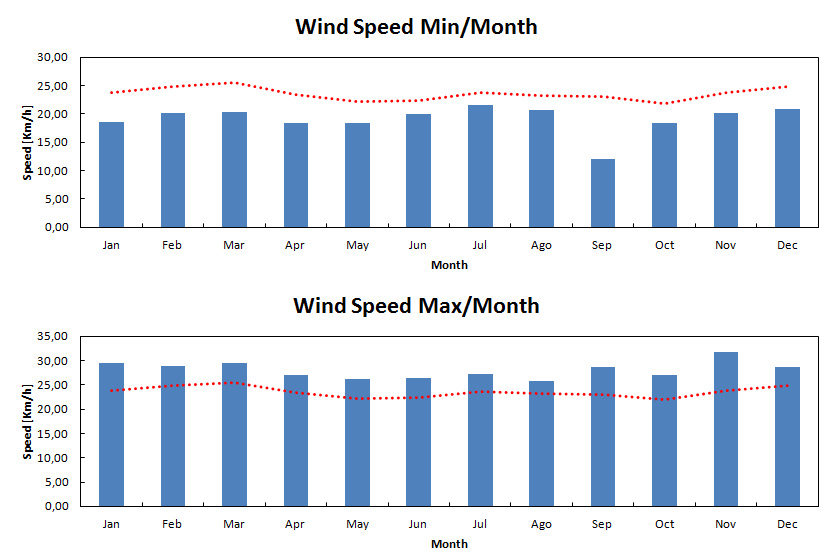}
\caption{Wind Speed (solid line) and Average Wind Speed Variations 1990-2014 (dotted line) }
\end{center}
\end{figure}
\clearpage
\begin{figure}
\includegraphics[width=1.3\textwidth{}]{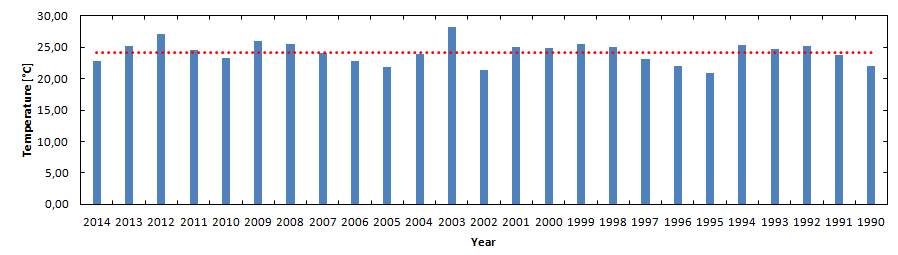}
\caption{August Temperature (solid line) and Average August Temperature 1990-2014 (dotted line) }
\end{figure}
\begin{figure}
\includegraphics[width=1.3\textwidth{}]{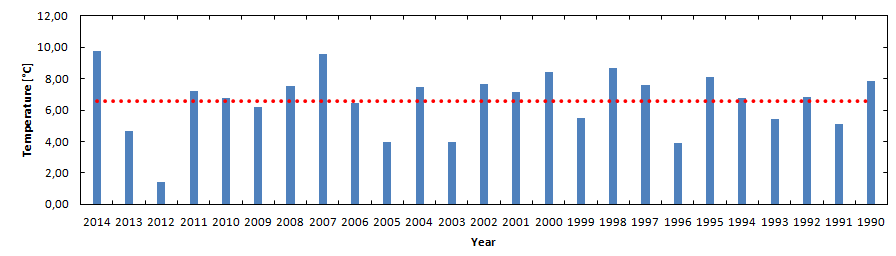}
\caption{February Temperature (solid line) and Average February Temperature 1990-2014 (dotted line)}
\end{figure}

\centering
\begin{figure}
\begin{center}
\includegraphics[width=1.3\textwidth{}]{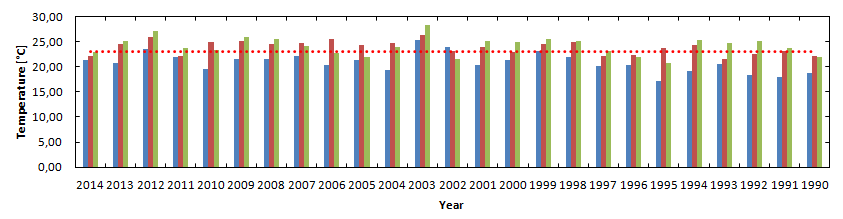}
\caption{Summer Temperature (solid line) and Average Temperature Variations 1990-2014 (dotted line). Jun: blu, Jul: red Aug: green}
\end{center}
\end{figure}
\begin{figure}
\begin{center}
\includegraphics[width=1.3\textwidth{}]{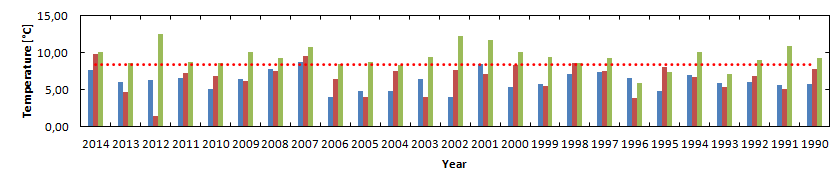}
\caption{Winner Temperature (solid line) and Average Temperature Variations 1990-2014 (dotted line) Jan: blu, Feb: red Mar: green}
\end{center}
\end{figure}
\begin{figure}
\caption{Total Rain (mm) by Month}
\includegraphics[width=0.55\textwidth{}]{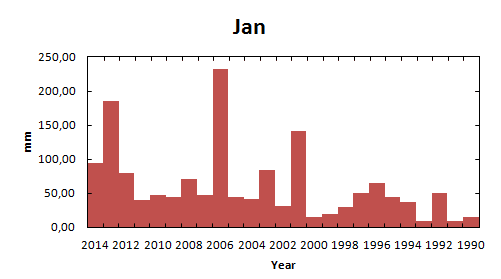} \quad 
\includegraphics[width=0.55\textwidth{}]{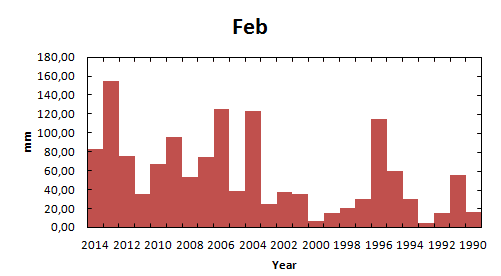} \quad
\includegraphics[width=0.55\textwidth{}]{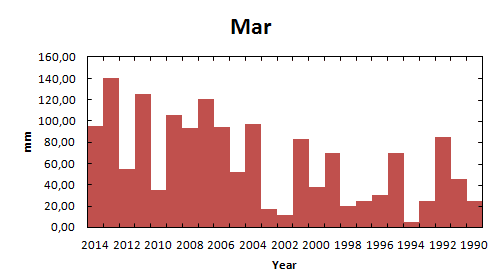} \quad
\includegraphics[width=0.60\textwidth{}]{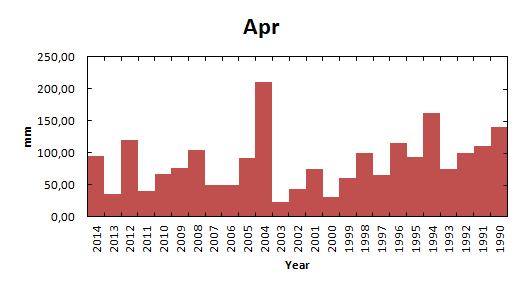} \quad
\includegraphics[width=0.55\textwidth{}]{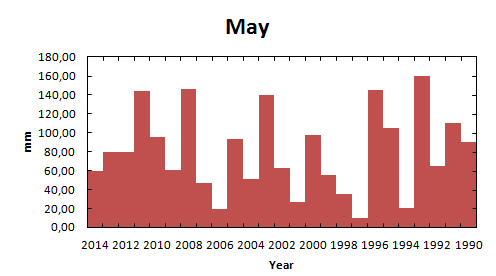} \quad 
\includegraphics[width=0.55\textwidth{}]{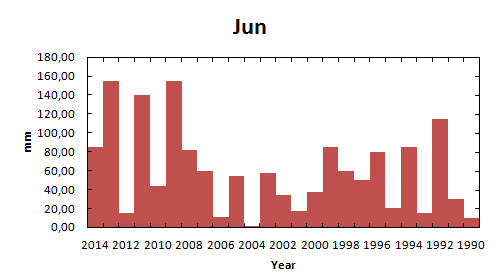} \quad
\includegraphics[width=0.55\textwidth{}]{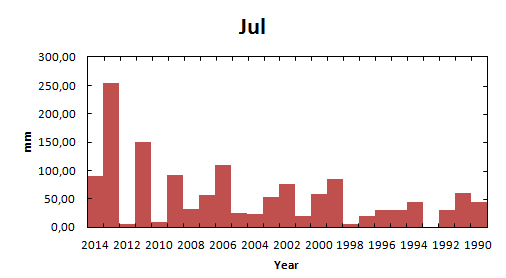} \quad
\includegraphics[width=0.55\textwidth{}]{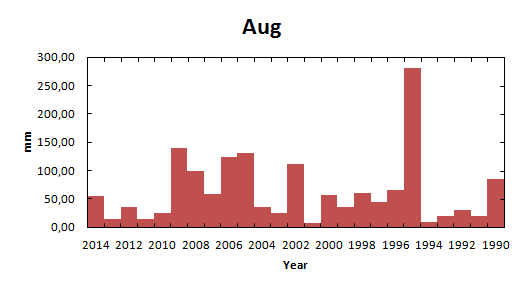} \quad
\includegraphics[width=0.55\textwidth{}]{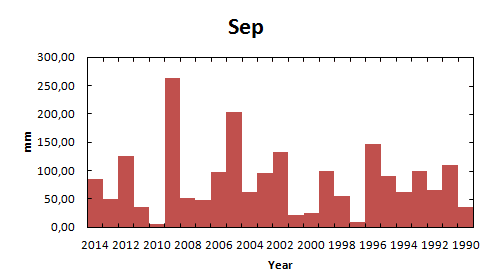} \quad 
\includegraphics[width=0.55\textwidth{}]{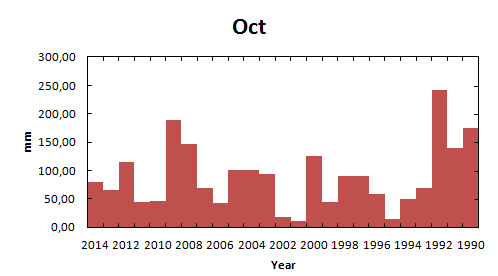} \quad
\includegraphics[width=0.55\textwidth{}]{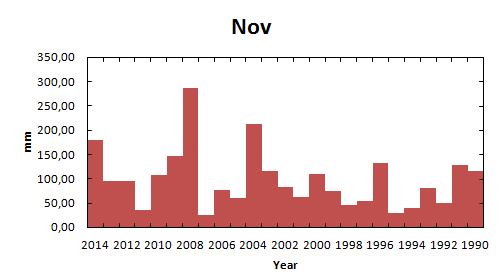} \quad
\includegraphics[width=0.55\textwidth{}]{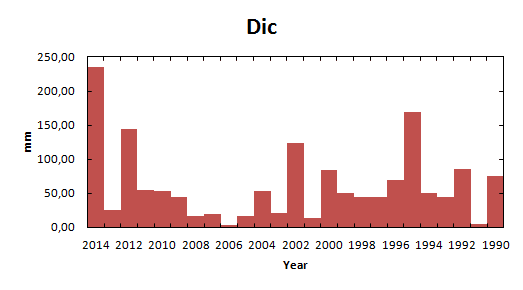} \\
\end{figure}
\begin{figure}
\caption{Solar Irradiation Map}
\includegraphics[width=1\textwidth{}]{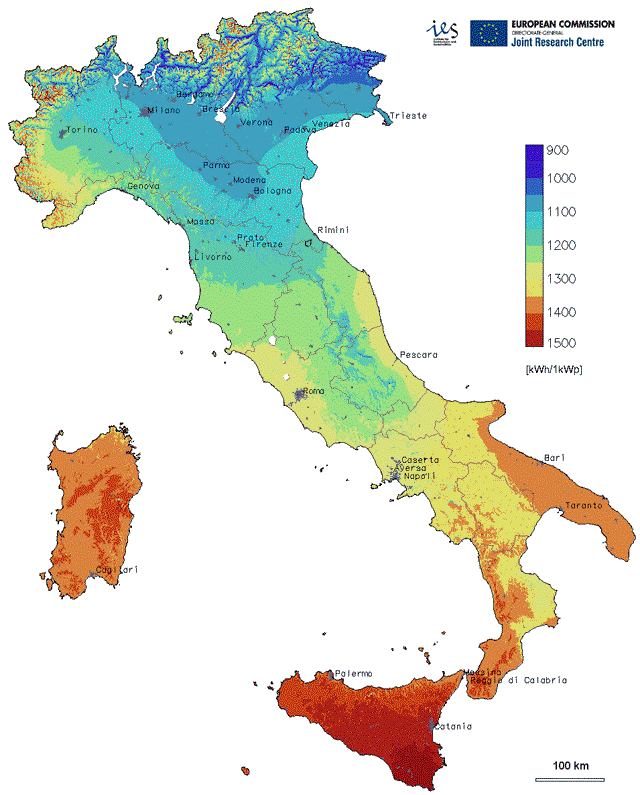}
\end{figure}
\begin{figure}
\includegraphics[width=0.9\textwidth{}]{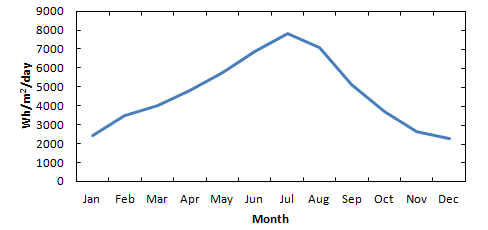}
\caption{DNI Direct Normal Irradiance (Wh/m$^2$/day)} 
\end{figure}
\clearpage 
\centering

\end{document}